\documentstyle[prl,aps,preprint,epsf]{revtex}
\firstfigfalse
\begin{document}
\title{ 
 On the Triviality of Textbook Quantum Electrodynamics} 
\vskip -1 truecm

\author{S. Kim}
\address{Department of Physics, Sejong University,
Seoul 143-747, Korea}

\author{John~B.~Kogut }
\address {Physics Department, University of Illinois at Urbana-Champaign,
Urbana, IL 61801-30}

\author{ Maria--Paola Lombardo}
\address{Istituto Nazionale di Fisica Nucleare,
         Sezione di Padova, e Gr. Coll. di Trento, Italy}

\date{\today}
\maketitle
\vskip -1 truecm

\begin{abstract}

By adding a small, irrelevant four fermi interaction to the action of lattice Quantum 
Electrodynamics (QED), the theory can be simulated with massless quarks in 
a vacuum free of lattice monopoles. This allows an ab initio high precision, 
controlled study of the existence of "textbook" Quantum Electrodynamics with 
several species of fermions. The lattice theory possesses a second order 
chiral phase transition which we show is logarithmically trivial. The 
logarithms of triviality, which modify mean field scaling laws, are 
pinpointed in several observables. The result supports Landau's contention 
that perturbative QED suffers from complete screening and would have a 
vanishing fine structure constant in the absence of a cutoff.
 
\end{abstract}

\pacs { 
12.38.Mh,
12.38.Gc,
11.15.Ha
}
\newpage

Attempts to solve conventional field theory problems by modern lattice 
gauge theory techniques run up against myriad problems. Two of
those are: 1. lattice field configurations may possess unphysical
artifacts of the finite short distance cutoff; and, 2. simulations with
realistic physical properties, such as (almost) vanishing fermion masses,
are almost impossible \cite{HMC},\cite{HMD}.

These problems can be solved in large part by adding a small, irrelevant
four fermi interaction to standard lattice actions
with staggered fermions.  The resulting lattice action can be simulated 
directly in the chiral limit (massless fermions) because an
auxiliary scalar field $\sigma$ (essentially the chiral 
condensate $<\bar\psi \psi>$) acts as a dynamical mass term for the
quarks insuring that the inversion of the Dirac operator will 
be successful and very fast. In addition, in the case of lattice QED
supplemented with a four fermi interaction, there is a second order
chiral transition where a continuum field theory may exist and the 
gauge field configurations for couplings near the transition
are free of lattice artifacts, such as monopoles and Dirac strings, etc.

Consider the 
$U(1)-$gauged Nambu Jona Lasinio (GNJL) model with four 
species of fermions. Because of the irrelevance of the pure
four fermi interaction, this model will make "textbook" QED
accessible and this paper will address the classic problem
of whether QED suffers from complete charge screening. Our
measurements will show that the theory is logarithmically trivial.

The Lagrangian for the continuum GNJL model is,

\begin{equation}
L  = \bar \psi (i\gamma \partial -e \gamma A - m) \psi -
\frac{1}{2}G^2 (\bar \psi \psi) ^2 -\frac {1}{4} F^2
\end{equation}

The Lagrangian has an electromagnetic interaction with continuous 
chiral invariance
($\psi \rightarrow  e^{i\alpha \tau \gamma_5} \psi$, where $\tau$
is the appropriate flavor matrix) and
a four fermi interaction with discrete ($Z_2$) 
chiral invariance
($\psi \rightarrow \gamma_5
\psi$). The mass term $m \bar\psi\psi$ breaks the chiral symmetries and will be set to zero.
The pure NJL model has been solved at large $N$ by
gap equation methods \cite{RWP}, and an accurate simulation study of it has been
presented \cite{Looking}.

To begin, we introduce an auxiliary random field $\sigma$ by adding
$-\frac{G^2}{2} ((\bar\psi\psi) -\frac{ \sigma}{G^2})^2$ to the
Lagrangian. This makes the Lagrangian a quadratic form in the fermion
field so it can be analyzed and simulated by conventional methods. 
The model is then discretized by using staggered fermions.
The lattice Action reads: 

\begin{equation}
S  =  \sum_{x,y} \bar\psi(x) (M_{xy} + D_{xy}) \psi(y) + 
  \frac {1}{2 G^2} \sum_{\tilde x} \sigma ^2 (\tilde x) +
  \frac{1}{2 e^2} \sum_{x,\mu,\nu} F^2_{\mu\nu}(x)
\end{equation}

\noindent
where 
\begin{eqnarray}
F_{\mu\nu}(x)& = &\theta_{\mu}(x) + \theta_{\nu}(x+\hat{\mu}) +
\theta_{-\mu}(x+\hat{\mu}+\hat{\nu}) + \theta_{-\nu}(x+\hat{\nu}) \\
M_{xy}& = & (m + \frac{1}{16} \sum_{<x,\tilde x>} \sigma( \tilde x))
\delta_{xy} \\
D_{xy} & = & \frac{1}{2} \sum_\mu \eta_{\mu} (x) (
 e^{i\theta_{\mu}(x)} \delta_{x+\hat{\mu}, y}
- e^{-i\theta_{\mu}(y)} \delta_{x-\hat{\mu}, y} )
\end{eqnarray}

In this formulation $\sigma$ is defined on the sites of the dual
lattice $\tilde x$ \cite{CER}, and the symbol $<x,\tilde x>$ denotes
the set of the 16 lattice sites surrounding the direct site $x$.
The factors $e^{\pm i\theta_\mu}$ are the gauge connections and
$\eta_\mu(x)$ are the staggered phases, the lattice analogs of the
Dirac matrices. $\psi$ is a staggered fermion 
field and  $m$ is the bare fermion mass, which will be set to 0.
Note that the lattice expression for $F_{\mu\nu}$ is non-compact
in the lattice field $\theta_{\mu}$, while the gauge field couples to the
fermion field through compact phase factors which guarantee local gauge
invariance.

The global discrete symmetry of the Action (2) reads:

\begin{eqnarray}
\psi(x) & \rightarrow & (-1)^{x1+x2+x3+x4} \psi(x) \\
\bar \psi(x) & \rightarrow & -\bar \psi (x) (-1)^{x1+x2+x3+x4} \\
\sigma & \rightarrow & - \sigma.
\end{eqnarray}
where $(-1)^{x1+x2+x3+x4}$ is the lattice representation of $\gamma_5$.

A rough estimate of the improved performance of the algorithm can be made. 
To invert the lattice Dirac operator having a minimum eigenvalue $\lambda_{min}$ 
to an accuracy $R$ (residual), takes a number of sweeps $N$ of a conjugate 
gradient algorithm which scales as $R \propto \exp(-a \lambda_{min} N)$. But
$\lambda_{min}$ is typically proportional to $\sigma + m$, where $\sigma$
is the average of the $\sigma(x)$ field over the configuration. If in a 
simulation $m$ is $0.01$ while $\sigma$ is $0.10$, then the new algorithm
is a full order of magnitude faster than the old. Our computer experiments
of QCD with four fermi interactions as well as GNJL have proven to be at 
least as good as this  \cite{KD}.

We studied this model using the Hybrid Molecular Dynamics
algorithm because this method can treat the number of fermion flavors as a continuous
variable. The standard staggered fermion algorithm produces eight 
flavors in the continuum limit. The Hybrid Molecular Dynamics algorithm
can take the square root of the fermion determinant and produce a
theory with four species of fermions in the continuum limit. This
is the theory we chose to study. The trick used in simulations of QCD
where one reduces the number of fermion species by a factor of two by 
placing lattice pseudo-fermion fields
only on even sites does not apply to the NJL term in this action. Since the
Hybrid Molecular Dynamics algorithm is not exact, we carefully 
monitored our simulations for systematic errors \cite{KKL}. 

Interesting limiting cases of the above Action are the pure $Z_2$ NJL
model ($e=0$), which has a phase transition at $G^2 \simeq 2$ \cite{Looking} and
the pure lattice QED ($G=0$) limit, whose chiral phase transition is
near $\beta_e \equiv 1/e^2 = .204$ for four flavors \cite{Kogut}, \cite{Az}.
The pure QED ($G=0$) model also has a monopole percolation transition
which is probably coincident with its chiral transition at $\beta_e = .204$ \cite{KW}.
Past simulations of this lattice model have led to contradictory results    
\cite{Az}, \cite{Goc}, \cite{Goc2}. 
Since the GNJL model can be simulated at $m=0$ for all couplings, the results reported here will 
be much more precise and decisive than those of the pure lattice QED ($G=0$) limit. 

We scanned the 2 dimensional parameter space ($\beta_e$,$G^2$) using the Hybrid Molecular Dynamics 
algorithm and measured the chiral condensate and monopole susceptibility as 
a function of $\beta_e$ and $G^2$. Recall that non-compact lattice QED possesses 
monopole excitations and Dirac strings \cite{HW}. We are particularly interested, however, in
simulating the model in regions of its parameter space where these topological 
excitations are non-critical so they do not contribute to the model's continuum limit.
We found that as we increased $G^2$ and 
moved off the $G=0$ axis, the peak of the monopole susceptibility shifted 
from $\beta_e = .204$ at $G=0$ to $\beta_e = .244$ at large $G$.
By contrast the chiral transition point shifted to a larger $\beta_e$ than the monopole percolation transition for a
given value of $G$
and became distinct from the monopole percolation point as soon as $G$ became 
nonzero. The movement of the monopole percolation peak in the ($\beta_e$,$G^2$) 
plane can be understood by noticing that $\sigma$ in the
Action plays the role of a site dependent mass term (Eq. 4). When the
fluctuations of the $\sigma$ field are not important, as is the case at
large $G$, the gauge field dynamics becomes equivalent to 
QED with a bare mass given by the constant $\sigma$ value. So, as $G^2$
increases and $\sigma$ grows, the theory approaches the large
$m$ limit of QED, i.e. quenched QED, which has a monopole
percolation transition at $\beta_e = .244$ \cite{HW}. This result was confirmed
quantitatively in the simulation.
In conclusion, the chiral transition line extends from ($\beta_e$,$G^2$) = $(.204,0)$
to $(\infty, \simeq 2.)$, while the monopole percolation line 
extends from $(.204, 0)$ to $(.244, \infty)$. The two transitions
only coincide at the "pure" QED point, $G=0$. Thus, the gauged NJL model makes it 
possible to study the triviality of conventional $U(1)$ gauge interactions 
\underline{without} topological excitations, an important physics problem which has bedeviled
field theory for decades. 
Landau originally concluded in the context of perturbation theory that
QED would be a free field theory because fermion vacuum polarization
would screen the electric charge completely. Alternatively, if one
renormalized the theory holding the renormalized charge fixed, then the
effective charge measured on a particular smaller length scale would diverge ( Landau's
ghost ). Landau's argument was originally made in the context of pure QED,
but its conclusion is not effected by adding irrelevant operators into
the Lagrangian since irrelevant operators do not effect the long distance,
physical content of the theory. In fact, in the context of lattice QED and
the GNJL model, the $G = 0$ lattice model is not "pure QED". Since the
discrete differences of the lattice formulation differ from continuum
derivatives, irrelevant operators distinguish the two field
theories. In the formulation studied here where the weak four fermi
term is explicitly introduced for technical reasons, we are always
addressing the same physical issues of the more familiar case of
"pure QED".

We have made accurate measurements on the chiral critical line for many choices of
couplings ($\beta_e$, $G^2$) and lattice sizes ranging from $8^4$ to $20^4$. Here
we shall discuss highlights of our data collected varying $\beta_e = 0.15-.30$ 
at fixed $G^2 = 1/4$ on a $16^4$ lattice and leave a more thorough presentation to another, lengthier
presentation. Finite size effects, efficiency and errors in the algorithm, measurement statistics 
and error analysis, as well as other scaling laws and critical indices will be dealt with elsewhere \cite{KKL}.

In Fig.1 we show the data for the chiral condensate $<\bar\psi \psi>$,
at fixed $G^2 = 1/4$ and variable $\beta_e$. The superb accuracy of this data, as
compared to typical lattice QCD data, will allow us to see, with a minimum of analysis, that the
chiral critical point is logarithmically trivial. 
Consider the  most conventional
fitting forms for this data. In mean field theory one predicts the equation of state 
$\beta_c-\beta_e = a \sigma^{1/\beta_{mag}}$, with
the critical index $\beta_{mag} = 1/2$. This scaling law is modified by logarithms in
trivial four dimensional models:
in the two component $\phi^4$ model, 
$\beta_c-\beta_e = a \sigma^{2}/\ln(b/\sigma)$ \cite{ID},
and in the $Z_2$ NJL model, $\beta_c-\beta_e = a \sigma^{2}\ln(b/\sigma)$ 
\cite{Looking}. In both of these simple models the interaction
strength falls to zero logarithmically as the cutoff is taken to infinity and this slow vanishing
of the interactions
causes the logarithmic effects in the equation of state for each model. It is interesting
that the logarithms enter differently in both equations of state, so we will be 
able to distinguish between $\phi^4$ triviality and NJL triviality. In fact $\phi^4$ triviality is almost
always assumed for "textbook" QED, but we shall find that NJL triviality is the actual answer.

The data shown in Fig.1 has been fit to a form which can accomodate either $\phi^4$ or NJL triviality:
$\beta_c-\beta_e = a \sigma^{2}(\ln(b/\sigma))^p$, where the parameter $p$, the critical point
$\beta_c$, the amplitude $a$ and the scale $b$ are determined by the 
fitting routine. For the scaling window of
gauge couplings $\beta_e$ between $.18$ and $.225$, we found 
the parameters $\beta_c = .2350(1)$, $a = 34.3(3.9)$, $\ln b = 1.55(10)$ and
$p = 1.00(8)$ with a confidence level of 34 percent. This is the fit shown in the figure. Since the
uncertainty in the power of the logarithm $p = 1.00(8)$ is so small, we have superb evidence for the
triviality of "textbook" QED. In fact, these simulations also measured topological observables for the system's
vacuum and we confirmed that monopoles and related objects were \underline{not} critical near
the chiral transition $\beta_c = .2350(1)$, $G^2 = 1/4$. ( We measured that 
the monopole percolation transition is very narrow and occurs at $\beta_e = .2175(25)$ for $G^2 = 1/4$
\cite{KKL}. )

\begin{figure}[htb]
\centerline{
\epsfxsize 5 in
\epsfysize 3 in
\epsfbox{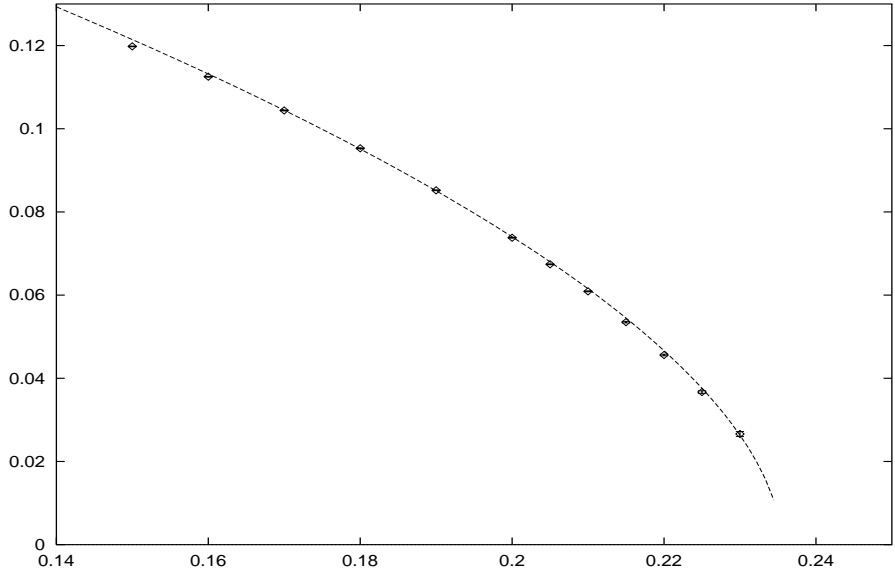}
}
\caption[]{$\sigma$ vs. $\beta_e$}
\end{figure}

The importance of the logarithm in the equation of state can be seen 
explicitly if we plot the data and the fit as shown in
Fig.2., $|\beta_c-\beta_e|/\sigma^2$ vs. $\ln(1/\sigma)$. If mean field theory were true, this plot
would be flat; if $\phi^4$ triviality applied, it would fall; and, if NJL triviality applied, it
would rise linearly. Fig. 2 shows that the third possibility is 
chosen decisively. The dashed line is the previous fit redrawn in this format.

\begin{figure}[htb]
\centerline{
\epsfxsize 5 in
\epsfysize 3 in
\epsfbox{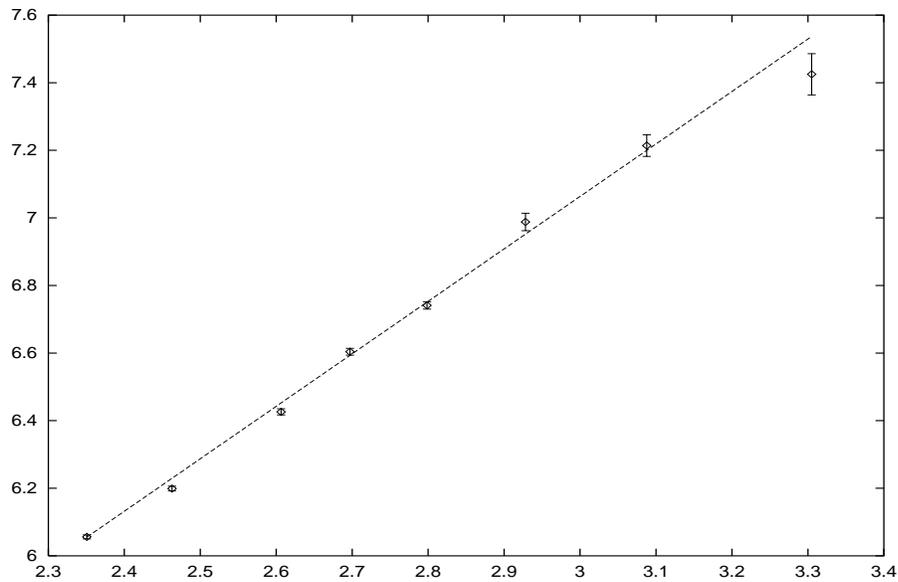}
}
\caption[]{$|\beta_c-\beta_e|/\sigma^2$ vs. $\ln(1/\sigma)$}
\end{figure}

Finally, in Fig.3 we show the inverse of the 
longitudinal susceptibility of the auxiliary field $\sigma$ 
at fixed $G^2 = 1/4$ and variable $\beta_e$. We plot the inverse of the
susceptibility rather than the susceptibility itself, because in mean field theory,
the singular piece of the longitudinal susceptibility $\chi$ 
diverges at the critical point $\beta_c$ as $\chi_+ = c_+ |t|^{-\gamma}$, 
$t \equiv (\beta_c-\beta_e)/\beta_c$, as $t$ approaches zero from above in the 
broken phase, and as $\chi_- = c_- |t|^{-\gamma}$ in the 
symmetric phase \cite{ID}. The critical index $\gamma$ is exactly unity in mean field theory.
As we see in the figure, the linear scaling law works well in the scaling 
windows on both sides of the critical point where
the inverse susceptibility vanishes.

\begin{figure}[htb]
\centerline{
\epsfxsize 5 in
\epsfysize 3 in
\epsfbox{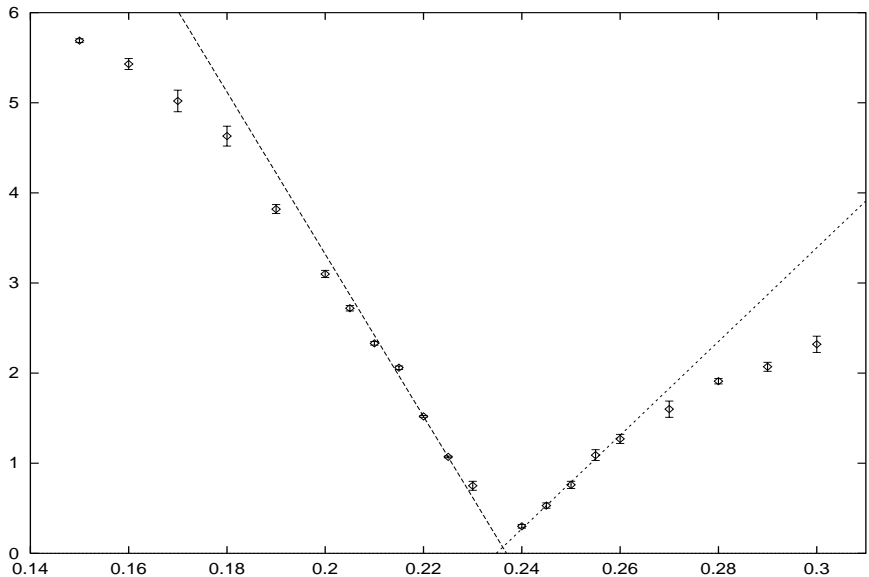}
}
\caption[]{Inverse Susceptibility vs. Coupling $\beta_e$}
\end{figure}

The plot picks out a critical point $\beta_c = .2358(5)$ and is consistent with 
the mean field value of the critical index $\gamma = 1.0$.

Another prediction of mean field theory is that
the universal amplitude ratio $c_-/c_+$ is exactly 2.0. However, in 
logarithmically trivial models $\gamma$ remains unity, but the amplitudes 
$c_+$ and $c_-$ develop weak logarithmic dependences \cite{ID}. In the 
two component $\phi^4$ model, $c_-/c_+ = 2 + \frac{2}{3}/\ln(\frac{b}{\sigma})$, while 
in the $Z_2$ NJL model, $c_-/c_+ = 2 - 1/\ln(\frac{b}{\sigma})$ 
\cite{Looking}, where the scale $b$ was determined in the order parameter fit. 
So, if $\phi^4$ triviality applied here we should find the amplitude
ratio slightly larger than $2$, and if NJL triviality applied we should find the amplitude
ratio slightly smaller than $2$. In fact, the constrained linear fits to 
the data shown in the figure produced the amplitude ratio $c_-/c_+ = 1.74(10)$. 
Since $\sigma$ varies from $.0953(1)$ to $.0367(2)$ over the $\beta_e$ range $.18$ - $.225$ of the
scaling window in the broken phase, the theoretical prediction of the NJL model states that
$c_-/c_+$ should range from $1.75$ to $1.79$. Again, the 
agreement between the simulation data and theory is very good.

    We have checked \cite{KKL} that the four fermi term is irrelevant
    as we get log-improved mean
    field theory in each of our runs at various $G^2 \ne 0$, ranging
    from $G^2 = 1/8$  to $G^2 = 1.$

Taken together Figs. 1-3 give a nicely consistent picture 
of the triviality of the four species
$U(1)$ gauged Nambu Jona Lasinio model with 
a $Z_2$ chiral group. We find no support for the 
approximate analytic schemes discussed in \cite{Az3}. 

It would be worthwhile to continue this work in several directions.
One could calculate the theory's renormalized couplings and their RG trajectories in the
chiral limit, extending the work of \cite{Goc2}. 
One could also simulate the model 
with the $Z_2$ chiral group replaced by a continuous group 
so the model would have Goldstone bosons even on a coarse lattice. 
Finally, it would be 
interesting to simulate 
compact QED with a small four fermi term and study the interplay of monopoles, charges and
chiral symmetry breaking. Since the $G = 0$ limit of the compact model 
is known to have a first order transition \cite{EDK},
generalizations of the action will be needed to
find a continuous transition where a continuum limit of the lattice theory might exist.

This work was partially supported by NSF under grant NSF-PHY96-05199. S. K. is
supported by the Korea Research Foundation. M.-P. L. wishes to thank the 
{\em ECT$^*$}, Trento, for hospitality during the final stages of this 
project.  The simulations were done at NPACI and NERSC. 



\begin{thebibliography}{99}
\bibitem{HMC} S. Duane, A.D. Kennedy, B.J. Pendleton and D. Roweth,
Phys. Lett. {\bf B195}, 216 (1987). 
\bibitem {HMD}
S. Duane and J.B. Kogut,  Phys. Rev. Lett. {\bf 55}, 2774 (1985). S. Gottlieb,
W. Liu, D. Toussaint, R.L. Renken and R.L. Sugar, Phys. Rev. {\bf D35},2531 (1987).
\bibitem{RWP}
B. Rosenstein, B. Warr and S. Park, Phys. Rep. {\bf 205}, 497 (1991).
\bibitem{Looking}
S. Kim, A. Koci$\acute{c}$ and J.B. Kogut, Nucl. Phys. {\bf B429}, 407 (1994).
\bibitem{CER}
Y. Cohen, S. Elitzur and E. Rabinovici, Nucl.
Phys. {\bf B220}, 102 (1983).
\bibitem{KD}
J.B. Kogut, J.-F. Lagae, and D.K. Sinclair, Phys. Rev. {\bf D58}, 34004 (1998). 
J.B. Kogut, and D.K. Sinclair, hep-lat/0005007.
\bibitem{KKL}
S. Kim, J.B. Kogut and M.-P. Lombardo, in preparation.
\bibitem{Kogut}
A. Koci$\acute{c}$, J.B. Kogut and K. C. Wang,
Nucl. Phys. {\bf B398}, 405 (1993).
\bibitem{Az}
V. Azcoiti, G. Di Carlo, A. Galante, A.F. Grillo, V. Laliena, and C.E. Piedrafita,
Phys. Lett. {\bf B353}, 279 (1995); {\bf B379}, 179 (1996).
\bibitem{KW}
J.B. Kogut and K.C.  Wang,  Phys.Rev. {\bf D53}, 1513 (1996).
\bibitem{Goc}
M. Gockeler, R. Horsley, P. Rakow, G. Schierholz and R. Sommer,
Nucl. Phys. {\bf B371}, 713(1992). 
\bibitem{Goc2}
M. Gockeler, R. Horsley, V. Linke, P. Rakow, G. Schierholz and H. Stuben, 
Phys. Rev. Lett. {\bf 80}, 4119 (1998). 
\bibitem{HW}
 S. Hands and R. Wensley, Phys. Rev. Lett. {\bf 63}, 2169 (1989).
\bibitem{ID}
C.~Itzykson and J.-M.~Drouffe, Statistical Field Theory (Cambridge
University Press, 1989.)
\bibitem{Az3}
V. Azcoiti, G. Di Carlo, A. Galante, A.F. Grillo, V. Laliena, C.E. Piedrafita
Phys.Lett. {\bf B355}, 270 (1995). 
\bibitem{EDK}
E. Dagotto and J.B. Kogut, Phys. Rev. Lett. {\bf 59}, 617 (1987).
\end{thebibliography}
\end{document}